\documentclass[fleqn,10pt]{wlscirep}
\usepackage[utf8]{inputenc}
\usepackage[T1]{fontenc}
\usepackage{mdframed}
\usepackage{xcolor}
\usepackage{array}

\title{Mobile Phone Location Data for Disasters: A Review from Natural Hazards and Epidemics}

\author[1]{Takahiro Yabe}
\author[2]{Nicholas K. W. Jones}
\author[1,3]{P. Suresh C. Rao}
\author[4,5]{Marta C. Gonzalez}
\author[1,*]{Satish V. Ukkusuri}

\affil[1]{Lyles School of Civil Engineering, Purdue University, 550 Stadium Mall Avenue, West Lafayette, Indiana 47907 USA}
\affil[2]{Global Facility for Disaster Reduction and Recovery, The World Bank, 1818 H Street, N.W. Washington, DC 20433 USA}
\affil[3]{Department of Agronomy, Purdue University, 550 Stadium Mall Avenue, West Lafayette, Indiana 47907 USA}
\affil[4]{Department of Civil and Environmental Engineering, UC Berkeley, 760 Davis Hall, University of California, Berkeley, California 94720, USA}
\affil[5]{Department of City and Regional Planning, UC Berkeley, 228 Bauer Wurster Hall, Berkeley, California 94720, USA}

\affil[*]{sukkusur@purdue.edu}


\begin{abstract}
Rapid urbanization and climate change trends are intertwined with complex interactions of various social, economic, and political factors. 
The increased trends of disaster risks have recently caused numerous events, ranging from unprecedented category 5 hurricanes in the Atlantic Ocean to the COVID-19 pandemic. 
While regions around the world face urgent demands to prepare for, respond to, and to recover from such disasters, large-scale location data collected from mobile phone devices have opened up novel approaches to tackle these challenges.
Mobile phone location data have enabled us to observe, estimate, and model human mobility dynamics at an unprecedented spatio-temporal granularity and scale. 
The COVID-19 pandemic has spurred the use of mobile phone location data for pandemic and disaster response. 
However, there is a lack of a comprehensive review that synthesizes the last decade of work leveraging mobile phone location data and case studies of natural hazards and epidemics. 
We address this gap by summarizing the existing work, and pointing promising areas and future challenges for using data to support disaster response and recovery. 
\end{abstract}
\begin{document}

\flushbottom
\maketitle
%
%
\thispagestyle{empty}


With population growth in many of the developing countries and concentration of resources and opportunities in urban areas, many cities around the world are experiencing rapid urbanization. 
The United Nations, Department of Economic and Social Affairs (UN DESA) estimates that by 2050, 68\% of the people in the world is projected to be living in cities, compared to 55\% in 2018 \cite{urbanization}.
In addition to rapid urbanization, continued anthropogenic emissions of greenhouse gases will cause further changes in the climate system, increasing the likelihood of severe and pervasive climate related hazards, including hurricanes, tropical cyclones, river floods, heat waves, and droughts \cite{ipcc}. 
Taken together, rapid urbanization and climate change, combined with complex interactions of various social, economic, and political factors, have increased and could further increase the risks of disasters across the globe.
For example, urbanization could lead to more population living in vulnerable locations to hazards, and more frequent disasters could widen the economic gap due to disproportionate impacts, which could then lead to political divide and instability. 
A ``disaster'' is a condition or event that leads to an unstable and dangerous situation for human society, and covers a wide range of shocks, including climate related hazards such as hurricanes, non-climate related natural hazards such as earthquakes and epidemics including COVID-19. 
Regions around the world need to urgently prepare for, respond to, and to recovery from these multitude of disasters for sustainable development. 

\begin{figure}[t]
    \centering
    \includegraphics[width=.85\textwidth]{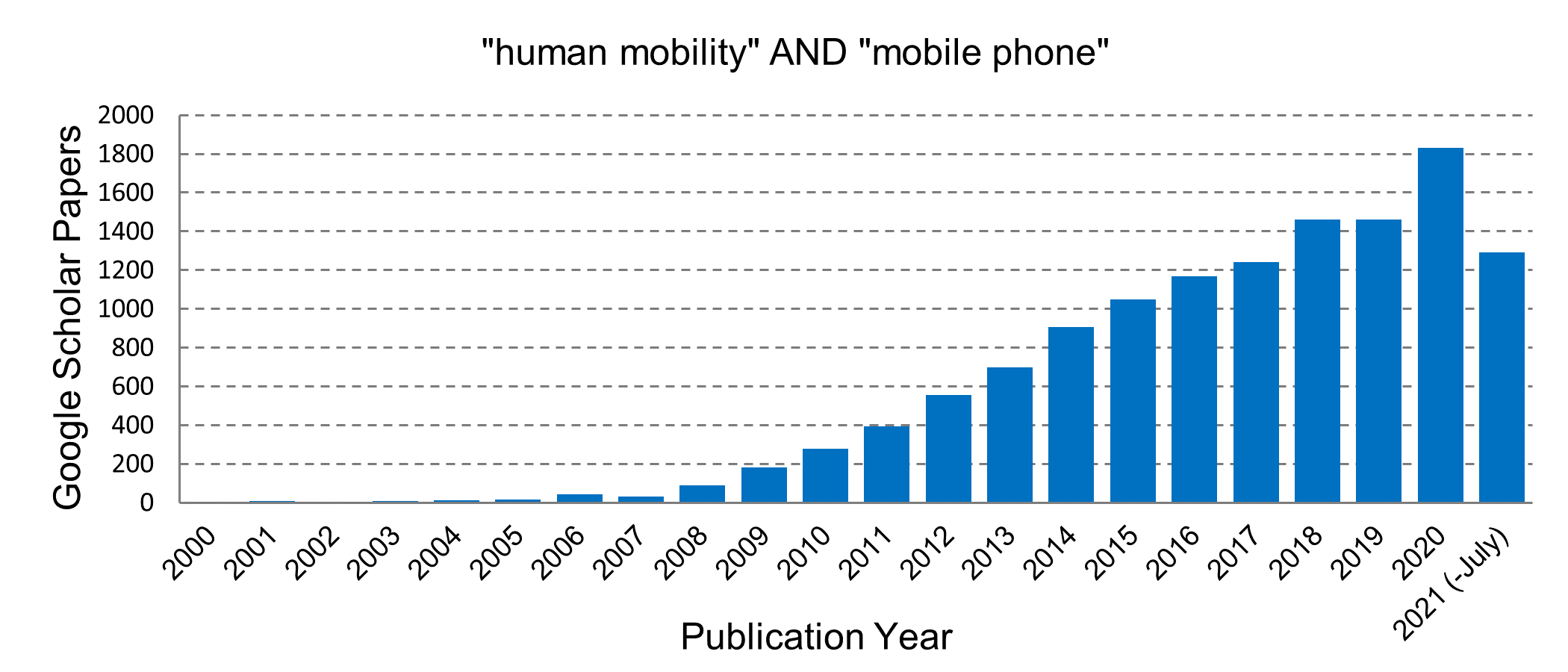}
    \caption{Number of research articles returned by searching ``human mobility'' and ``mobile phone'' in Google Scholar by year. Research has substantially increased over the years and was further spurred in 2020 due to the COVID-19 pandemic. The count for 2021 was computed on July 12th, 2021.}
    \label{fig:number}
\end{figure}

The pervasiveness of mobile devices (mobile phones, smartphones) across the globe has opened up massive opportunities to collect large-scale location data from individual users at an unprecedented scale compared to previous approaches (see Figure \ref{fig:number} for number of publications on human mobility and mobile phone data).
Human mobility (for a review, see Barbosa et al. \cite{barbosa2018human}) is a critical component to understanding various disaster events. 
Large scale natural hazards cause severe damage to housing structures and infrastructure systems, triggering mass evacuation, displacement, and migration from affected areas. 
For agencies who aim to aid those who fled their homes with essential services and supplies, locations of such movement destinations serve as crucial input information.  
Infectious diseases are by definition, transmitted between humans. 
Understanding the inter-regional mobility flows could assist epidemiologists predict the outbreak of the disease. 
Mobile phone location data are pertinent for responding to and recovering from such disaster events. 

Prior to the availability of mobile phone location data, household surveys have been the primary source of information on understanding human mobility. 
Household surveys, compared to mobile phone location data, are advantageous in collecting detailed information about respondents' socio-demographic and economic characteristics, and knowing the reasoning of why the respondents behaved in a certain manner.  
Mobile phone location data, despite its drawbacks in data governance and quality uncertainties (discussed in Section \ref{challenges} in detail), is able to provide us with location information of a massive number of samples (often millions), a rapid manner (minimum a few days; e.g., \cite{wilson2016rapid}), at a high frequency (e.g., around 50 data points each day), longitudinal time frame (e.g., 6 months before and after the disaster event \cite{yabe2020understanding}), and high spatial granularity ($\sim$100 meters in spatial error). 
More recently, the coronavirus disease (COVID-19) pandemic has spurred and accelerated the use of mobile phone location data for pandemic disaster response \cite{oliver2020mobile}. 
The attention and interest towards mobile phone location data from government agencies, researchers, and the public, has never been higher. 

Despite such interest in the analysis of mobile phone location data for disaster management, currently we lack a comprehensive review of literature that synthesizes the progress that has been made in the past decade. 
Cinnamon et al. \cite{cinnamon2016evidence} reviews the progress using mobile phone call detail record (CDR) data.
Yu et al. \cite{yu2018big} and Akter et al. \cite{akter2019big} review the usage and applications of various types of novel big data, including social media data and satellite image data.
Wang et al. \cite{wang2020using} reviews the usage of various mobile phone location data (including smartphone GPS data), and is the most recent and closest to our review. 
However, with the spread of COVID-19, more types of mobile phone location data has become increasingly available in research. 
Pressing social and technical issues around mobile phone location data, including the opaqueness of the data generative process and data governance, comparisons between different types of data (e.g., CDR vs GPS from location intelligence firms vs GPS from major tech firms), applications in COVID-19 response, and recent progress in collaborations between academia, government agencies, and the industry, are important topics that need to be reviewed for further progress in this area. 

The increasing use of mobile phone location data in disaster management and social good, recently spurred by global efforts in COVID-19 response, has highlighted the usefulness of these datasets for assisting response and recovery \cite{oliver2020mobile}. 
However, at the same time, concerns about personal privacy, data governance, and potential malicious uses of mobile phone location data have been raised as well \cite{nyt}.
To organize and understand what can be achieved using mobile phone location data for disaster management, and also its limitations, as well as methodological, societal, and data-related issues, this article conducts a comprehensive and interdisciplinary literature review on efforts that have used mobile phone location data for disaster management. 
This review will not cover, however, the other types of data that are more frequently being used in disaster management, including social media data (for a review, see e.g., Muniz et al. \cite{muniz2020social}, Kryvasheyeu et al. \cite{kryvasheyeu2016rapid}) and satellite imagery data (for a review, see e.g., \cite{joyce2009review}). 
In Section 2, we review the typology of mobile phone location data, Section 3 covers the scientific progress, applications, and case studies in natural hazards and epidemics. 
Section 4 and 5 discusses and concludes with opportunities and future challenges of using mobile phone location data for disaster response and recovery.

\section*{Types of Mobile Phone Location Data}

Mobile phone location data can be classified into three main categories: mobile phone call detail records (CDR), smartphone GPS location data collected by location intelligence companies, and smartphone GPS location data collected and processed by major tech companies.
Table \ref{tab:1data} organizes how they are collected, the pros, cons, and examples of providers for each dataset. 

\subsection*{Mobile Phone Call Detail Records (CDR)}

During the last decade, mobile phone call detail records (CDR) have become one of the primary data sources for analyzing human mobility patterns on the urban scale \cite{calabrese2011estimating}.
Call detail records typically contain the unique ID of the user, timestamp, and location information of the observed cell phone tower. 
Note that unlike smartphone GPS data introduced later, the location information of CDRs are not the actual location of the user, thus contains typically around couple 100 meters to several kilometers in the rural areas where cell phone towers are sparsely located.  
Using large-scale datasets of CDR data, a seminal paper by Gonzalez et al. unraveled the basic laws of human mobility patterns \cite{gonzalez2008understanding}. 
Several more papers have used CDR data to understand spatio-temporal patterns of urban human mobility, routine behavior, and their predictability (e.g., \cite{hasan2013spatiotemporal,song2010limits}). 
Moreover, human activity patterns and land use patterns have been studied using CDR data (e.g., \cite{pei2014new}).
In addition to understanding human behavioral laws, such data has enabled us to obtain dynamic and spatially detailed estimations of population distributions (e.g., \cite{deville2014dynamic}), social integration and segregation of mobility (e.g., \cite{phillips2019social}), and macroscopic migration patterns (e.g., \cite{blumenstock2012inferring}). 
Moreover, these datasets have been applied to solve various urban problems such as preventing disease spread \cite{finger2016mobile,wesolowski2012quantifying,bengtsson2015using}, estimating traffic flow (e.g., \cite{iqbal2014development,alexander2015origin}), and estimating socioeconomic statistics (e.g., \cite{xu2018human}) and impacts of shocks \cite{toole2015tracking} (for a full review, see \cite{blondel2015survey,jiang2013review}). 

\begin{table}[t]
    \centering
    \begin{tabular}{m{3cm} m{4cm} m{4.5cm} m{3cm}}
    \toprule
    Data type & Description & Pros and Cons & Providers (e.g.) \\ 
    \midrule
    Mobile phone call detail records (CDR) & Location information of cell phone towers when users make calls or text messages & (+) substantial coverage of the population (-) Low spatial and temporal resolution compared to GPS datasets & NCell, Orange, Vodafone, Turkcell \\
    \midrule
    Smartphone GPS location data (Location Intelligence firms) & GPS data collected and aggregated from several third party smartphone applications & (+) precise location information of users (-) No transparency in data generation process; covers a small sample of population compared to CDR; available for fewer countries & Cuebiq, Veraset, Safegraph, Unacast \\
    \midrule
    Smartphone GPS location data (Major Tech firms) & GPS data collected and aggregated from their own platforms & (+) Available in standardized formats across multiple countries and across time (-) Outputs restricted to selected metrics produced by the tech firms & Google, Facebook, Apple, Yahoo Japan \\
    \bottomrule
    \end{tabular}
    \caption{Brief descriptions and applications of the four novel types of data: mobile phone location data, social media data, web search query data, and satellite imagery night time light data.}
    \label{tab:1data}
\end{table}

\subsection*{Smartphone GPS Location Data from Location Intelligence Firms}

More recently, we have seen an increase in the availability of mobile phone GPS location datasets collected by location intelligence companies, such as Cuebiq (\url{https://www.cuebiq.com/}), Unacast (\url{https://www.unacast.com/}), and Safegraph (\url{https://www.safegraph.com/}). 
Location intelligence companies collect location data (e.g., GPS data) from third-party data partners such as mobile location-based application developers.
Typically for each data point, a user identifier, timestamp of observation, and the longitude and latitude information are included in the dataset. 
More recently, these firms have started provided more aggregate (e.g., aggregated for each point-of-interest) data to preserve the privacy of the users. 
Compared to CDR, GPS logs have higher spatial preciseness, and moreover, higher observation frequency, allowing us to understand mobility patterns in more detail. 
However, often the specific sources of the location data nor the process in which the data are collected and combined from several application services are undisclosed to the users.  
Therefore, using such data requires a rigorous analysis of checking the representativeness of the mobile phone location dataset.

\subsection*{Smartphone Location Data from Major Tech Firms}

Similar to the smartphone GPS location data collected by location intelligence firms, major tech firms such as Facebook, Google, and Apple, also collect GPS location data from their users. 
The major difference in the data generative process is that these major tech firms use data collected from their own platform, not by third party services. 
Often, these data are provided in a pre-processed form, aggregated by both time and space.
Facebook, through its ``Data for Good'' program, provides various types of location information products to researchers, agencies, and non-profits (\url{https://dataforgood.fb.com}).
In particular, the ``Facebook Disaster Maps'' provides detailed density maps of the population density and movement patterns before, during, and after disaster events.
The data is temporally aggregated (usually every 24 hours), spatially aggregated (usually into 360,000 square meter tiles), and spatially smoothed, to anonymize and protect the users' privacy \cite{maas2019facebook,jia2020patterns}.
The Maps have been utilized by many significant nonprofit organizations and international agencies in disaster response, including the International Federation of the Red Cross, the World Food Programme, the United Nations Children’s Fund (UNICEF), NetHope, Direct Relief, and others.


\section*{Applications and Methodologies}


\subsection*{Natural Hazards}

Recently, mobile phone data has been utilized in many applications for disaster response and recovery, given its high spatial and temporal granularity, scalability to analyze millions of individuals' mobility, and increasing availability. 
In this section, the studies using mobile phone data for natural hazard response and recovery are categorized into 3 categories of applications: population displacement and evacuation modeling, longer-term recovery analysis, and inverse inference of damages to the built environment.
The required inputs, methodologies, obtained outputs, and case studies are presented for each application. 

\subsubsection*{Population Displacement and Evacuation Modeling}

The most widely studied applications of mobile phone location data in disaster response and recovery is to estimate the population displacement and evacuation dynamics after disasters. 
In their seminal paper, Lu et al. used CDR to study the predictability of displacement mobility patterns after the Haiti Earthquake in 2010 \cite{lu2012predictability}. 
Using data collected from 1.9 million mobile phone users during the period from 42 days before to 341 days after the shock, the study estimated that 23\% of the population in Port-au-Prince had been displaced due to the earthquake. 
Despite the substantial displacement, they also found that the destinations of the displaced people were highly correlated with their pre-earthquake mobility patterns.
This finding shed light on the possibility of predicting post-disaster mobility patterns, and had significant implications on relief operations including the pre-positioning of distribution centers \cite{yushimito2012voronoi} and evacuation shelters. 
Another seminal disaster event that highlighted the use of mobile phone location data was the Gorkha Earthquake (intensity of 7.8Mw) which struck Nepal in 2015 \cite{kargel2016geomorphic}.
Wilson et al. rapidly analyzed the displacement movements of 12 million de-identified mobile phone users after the earthquake within nine days from the event \cite{wilson2016rapid}. 
It was estimated that over 390,000 people left the Kathmandu Valley after the earthquake.
These results were released as a report with the United Nations Office for the Coordination of Humanitarian Affairs (UN OCHA) and a range of relief agencies. 
This effort by Flowminder, a non-profit foundation for analyzing mobile phone location datasets, was the first significant practical use-case of large scale mobile phone location data in disaster relief and response \cite{flowminder}. 

\begin{figure}[t]
\centering
\includegraphics[width=.85\linewidth]{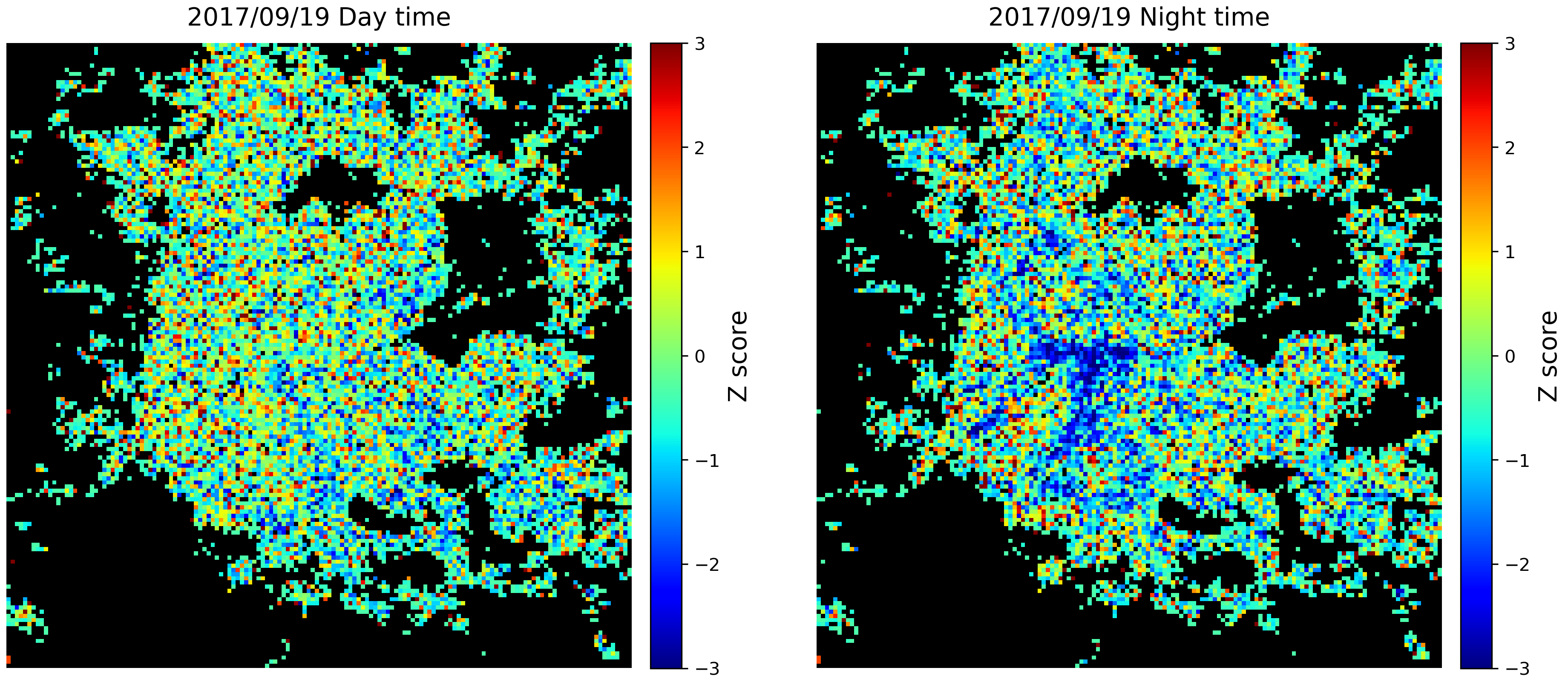}
\caption{\textbf{Population displacement after the Puebla Earthquake in Mexico City.} Anomaly score ($z$ score; number of standard deviations more/less than the pre-earthquake mean) of population density during the day (left) and night (right) on September 19th, 2017 in Mexico City. Significant displacement is observed during the night time of the day of the earthquake (Source: \cite{gfdrrworkingpaper})}
\label{fig:evac}
\end{figure}

Following the aforementioned two seminal works after the Haiti Earthquake and Gorkha Earthquake, several studies have developed methods to estimate population displacement and post-disaster evacuation patterns using mobile phone location data. 
A general framework for the spatio-temporal detection of behavioral anomalies using mobile phone data was proposed by Dobra et al. \cite{dobra2015spatiotemporal}. 
Using smartphone location data from before and after disasters, population displacement can be quantified by measuring the anomaly score (z-score; the number of standard deviations more or less from the mean population on a typical day) of the daytime and nighttime population in highly granular (1km x 1km) grid cells, as shown in Figure \ref{fig:evac} \cite{gfdrrworkingpaper}.
During the night time after the earthquake, blue-colored clusters with z scores below -2, indicating a likelihood of less than 1\% on a typical day, can be observed in central Mexico City, showing significant decrease in nighttime population.
Yabe et al. used smartphone GPS location data collected by Yahoo Japan Corporation to analyze the evacuation rates after five earthquake events in Japan \cite{yabe2018cross}. 
Cross-comparative analysis of five earthquakes and over 100 affected communities revealed similar relationships between evacuation rates and seismic intensity levels, where evacuation rates significantly increased in communities that experienced magnitudes above 5.5.
Several computational frameworks have been proposed to estimate the spatial patterns of evacuation destinations and hotspot locations using anomaly detection techniques on large-scale mobility data \cite{yabe2016framework}. 
Just after the Kumamoto Earthquake in April 2016, population distribution and evacuation hotspot maps were produced jointly by researchers at the University of Tokyo and Yahoo Japan Research, and were delivered to city governments for relief and response \cite{yahoobigdata}. 
Duan et al. studied the evacuation patterns after a train collision incident in China using mobile phone location data, identifying a two-stage evacuation process, and also behavioral changes in commuters' travel route choices \cite{duan2017understanding}. 
Ghurye et al. study the displacement patterns after the Rwanda Flood in 2012 using Markov Chain models and CDR \cite{ghurye2016framework}. 
The study compares the observed human behavior during a disaster with the behavior expected under normal circumstances to understand the causal effects of the disaster event. 
Yin et al. combined mobile phone location data with agent based simulations (which are widely used in evacuation analysis; e.g., \cite{ukkusuri2017rescue}) to improve the estimation accuracy of evacuation movement, proposing a hybrid approach \cite{yin2020improving}. 

More computational approaches using data assimilation techniques have been explored for online, near real-time predictions of post-disaster mobility patterns. 
Song et al. proposed a mobility prediction model based on a Hidden Markov Modeling framework, and tested its validity using data collected from 1.6 million mobile phone users in Japan before, during, and after the Great East Japan Earthquake in 2011 \cite{song2014prediction}. 
Sudo et al. developed a Bayesian data assimilation framework by combining the particle filter and Earth Mover's Distance algorithms, that updates the urban-scale agent based mobility simulation in an online manner using spatially aggregate mobile phone location data provided in real time \cite{sudo2016particle,sekimoto2016real}.
Several online algorithms have been proposed since these seminal works, including CityMomentum \cite{fan2015citymomentum} that uses a mixture of multiple random Markov chains, CityCoupling \cite{fancoupling} that aims to perform cross-city predictions, and inverse reinforcement learning approaches that attempt to learn the behavioral patterns of human mobility during disasters from large scale data \cite{pang2018replicating,pang2020intercity}. 
Although these computational, online approaches are shown to be effective in experimental and post-hoc settings, none have been utilized in real-time after real-world disaster events. 

\begin{mdframed}[backgroundcolor=gray!20] 
\textbf{\textit{Policy applications:}} Evacuation and displacement estimation could be used for making various policy decisions during the response and preparation stages of the disaster risk management cycle, including quick identification of post-disaster needs, planning of emergency supply distribution networks, and pre-positioning of evacuation shelters and supplies. 
\end{mdframed}

\subsubsection*{Longer-term Analysis: Migration and Recovery}

\begin{figure}[t]
\centering
\begin{minipage}{0.32\textwidth}
    \centering
  \includegraphics[width=\textwidth]{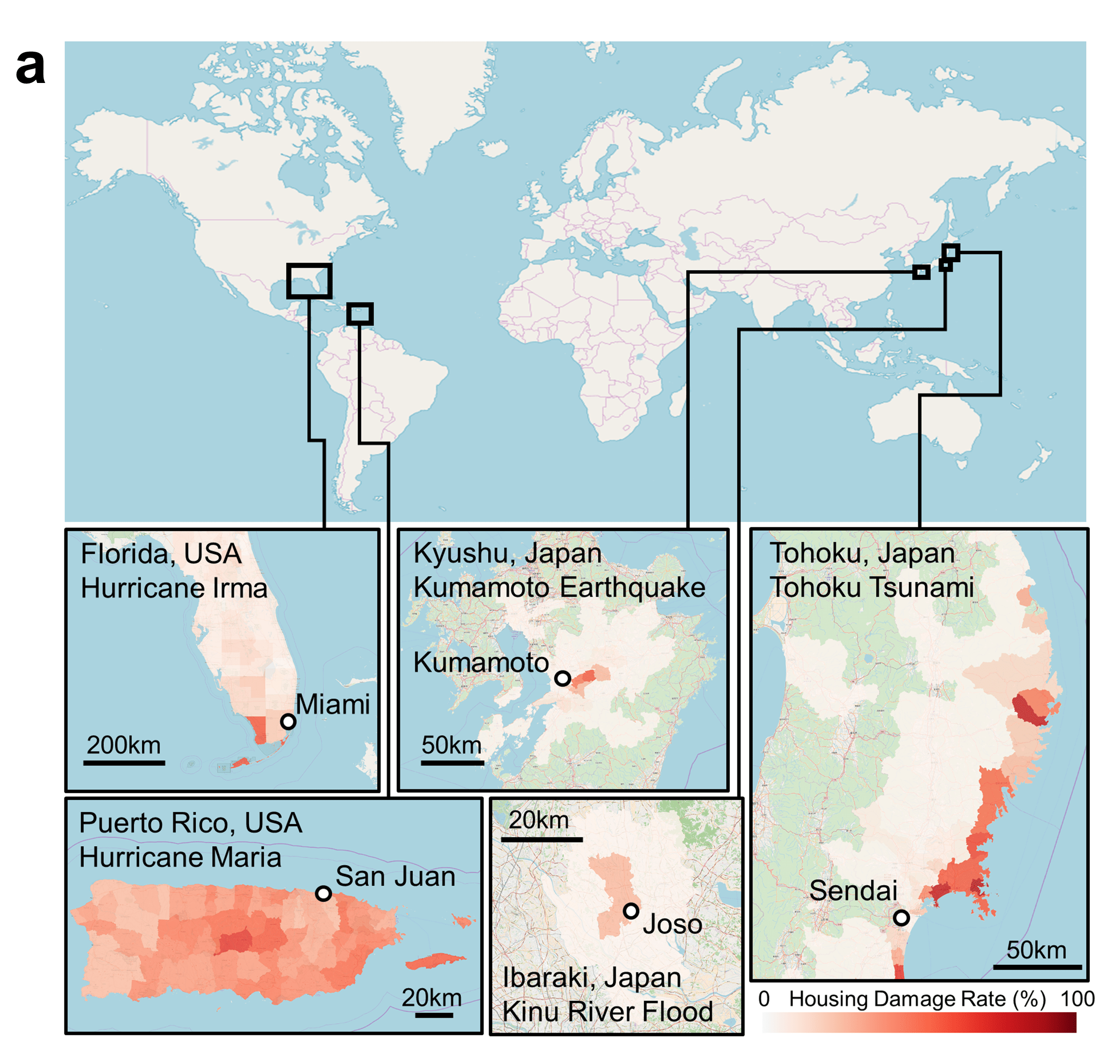}
  \label{fig:locations}
  \end{minipage}
\begin{minipage}{0.67\textwidth}
\centering
\includegraphics[width=\textwidth]{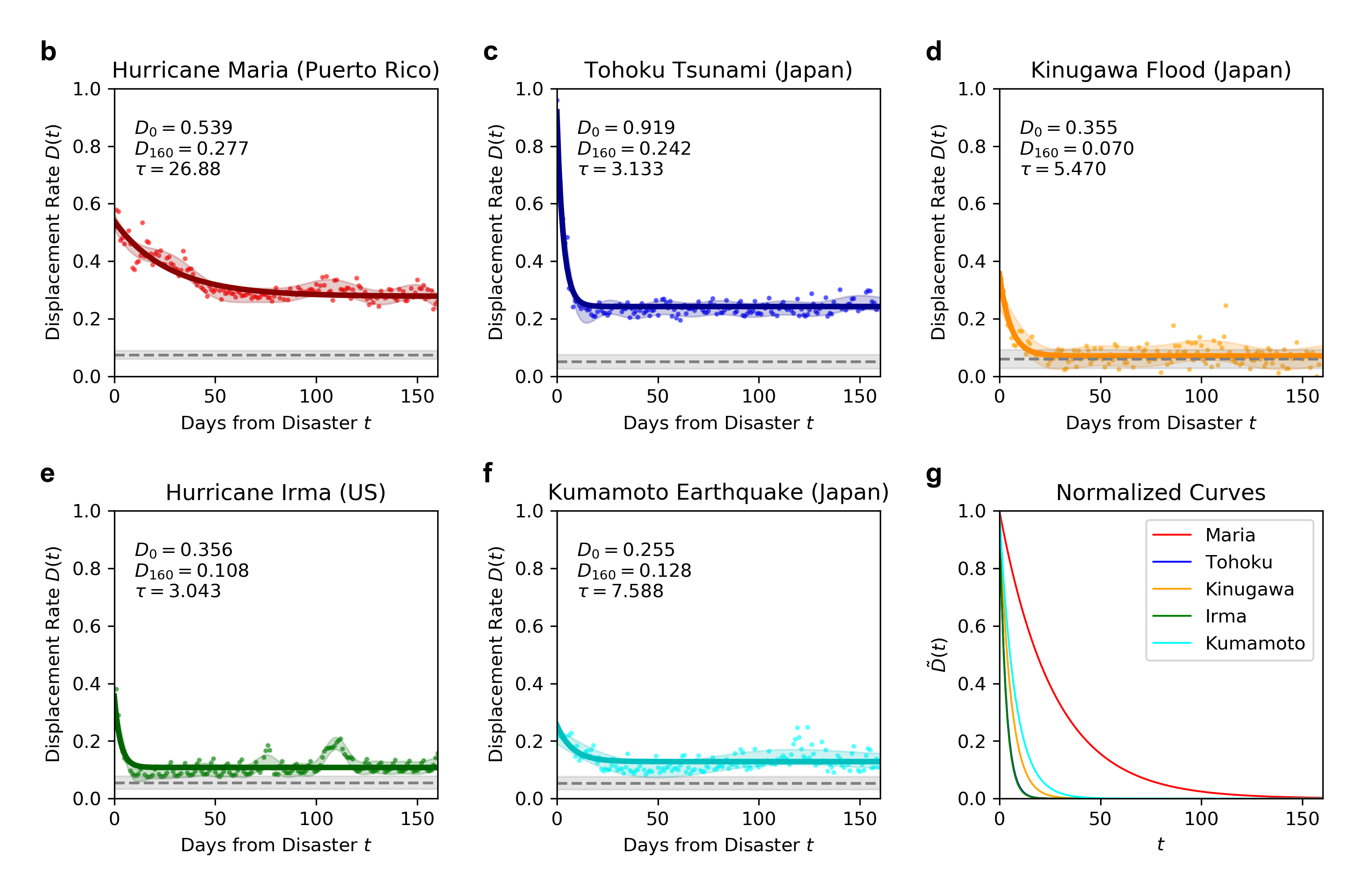}
\label{fig:ps}
\end{minipage} 
\caption{
\textbf{Similarity of macroscopic population recovery patterns across the five disasters.}  
\textbf{a.} 
Location, spatial scale, and severity of disasters that were studied. Red colors indicate the percentages of houses that were severely damaged in each community.
\textbf{b-f.} 
Macroscopic population recovery patterns after each disaster. Raw observations of displacement rates were denoised using Gaussian Process Regression and were then fitted with a negative exponential function. $D_0$, $D_{160}$ and $\tau$ denote the displacement rates on day 0, day 160, and recovery time parameter of each fitted negative exponential function. Black horizontal dashed line shows average displacement rates observed before the disaster.
\textbf{g.} 
Normalized population recovery patterns all follow an exponential decay. (Source: \cite{yabe2020understanding})
}
\label{fig:univpatterns}
\end{figure}

One advantage of mobile phone location data is the ability to track the movements of users over a long period of time (several months $\sim$ year) with high frequency (e.g., hourly $\sim$ daily), which are extremely difficult to perform using household survey data. 
Therefore, in the normal setting, there have been attempts to use mobile phone location data to estimate population migration dynamics \cite{hankaew2019inferring,lai2019exploring,blumenstock2012inferring}. 
In the disaster setting, Lu et al. studied the migration patterns in regions stressed by climate shocks in Bangladesh using CDR \cite{lu2016unveiling}. 
In addition to analyzing the short term human mobility patterns after Cyclone events (hours $\sim$ weeks), the study quantifies the incidence, direction, duration and seasonality of migration in Bangladesh.
Acosta et al. quantified the migration dynamics from Puerto Rico after Hurricane maria using mobile phone, showing a shift from rural to urban areas after the disaster \cite{acosta2020quantifying}. 
Yabe et al. studied the population displacement and recovery patterns after five disaster events, including Hurricanes Maria and Irma, earthquakes, floods, and tsunami using mobile phone GPS datasets from Japan and the US (Figure \ref{fig:univpatterns}) \cite{yabe2020understanding}. 
Cross-comparative analysis of five major disasters revealed general exponential decay patterns of population recovery, and predictability using a small number of key socio-economic factors including population density, infrastructure recovery patterns, wealth indexes, and spatial network patterns. 
Marzuoli et al. used mobile phone data to analyze the recovery dynamics of residents in South Texas after Hurricane Harvey \cite{marzuoli2018data}. 
The study provided detailed statistics of population movement and origin destination patterns for different zipcodes in Texas. 
In addition, the role of social networks \cite{sadri2018role,yabe2019mobile}, hedonic behavior \cite{jia2017role}, and post-disaster spatial segregation \cite{yabe2020effects} have been tested using mobile phone location data after disasters. 
Apart from population recovery and migration analysis, mobile phone location data has been used to quantify the recovery of business firms as well, using foot traffic counts as a proxy for revenue \cite{yabe2020quantifying}. 
Regional and sector differences in disaster impacts have been quantified via a Bayesian structural time series framework, using data collected from over 200 business entities in Puerto Rico before and after Hurricanes Irma and Maria. 
Although mobile phone location data provide significant advantages in analyzing longer-term phenomena (e.g., migration and recovery after disaster events) compared to household survey data, most studies focus on shorter term displacement and evacuation analysis, leaving substantial room for research in understanding the long term recovery and resilience of urban and rural areas to disasters. 

\begin{mdframed}[backgroundcolor=gray!20] 
\textbf{\textit{Policy applications:}} Analysis of migration and recovery estimation could be used for developing policies that focus on longer term recovery and mitigation of hazards. 
Such tasks include longer term infrastructure investment plans for building-back-better from disasters, strategies for harnessing community social capital for community resilience, and planning of urban land use master plans to prepare for longer term migration dynamics. 
\end{mdframed}

\subsubsection*{Inverse Inference of Damage to the Built Environment}

The studies introduced in the previous two subsections studied the anomalies in human mobility patterns disrupted by shocks (e.g., hurricanes, earthquakes, tsunami) inflicted to the built environment. 
However, several studies have approached the problem in an inverse manner, by using anomalies observed in the mobile phone location data and human mobility dynamics to inversely estimate the damage to and recovery of the built environment, which have traditionally been estimated using hazard simulations and structural mechanics (e.g., \cite{baker2015efficient}). 
Andrade et al. propose a novel metric ``reach score'' that quantifies the amount of movement of mobile phone users, and finds that the reach score has significant correlation with the damage inflicted to infrastructure systems by the earthquake at the canton level in Ecuador \cite{andrade2018risc}. 
Pastor-Escuredo et al. show that by analyzing the anomalous patterns in mobile phone communications, we are able to conduct infrastructure impact assessment due to flooding events, using retrospective data collected from a flood in Mexico \cite{pastor2014flooding}. 
Finally, Yabe et al. propose a machine learning algorithm that combines mobile phone location data with terrain information to conduct a rapid and accurate estimate of the inundated areas during a flood event \cite{yabeflood}. 
These studies show the potential of using mobile phone location data to infer the abnormal states of the built environment. 
Mobile phone location data has several advantages compared to conventional methods in data quality, including satellite imagery which are often observed sparse in time (e.g., once a day at most), and social media data which are more sparsely observed. 
While the application potential of these studies are promising, we lack comprehensive analysis of its real-time feasibility and accuracy under different types of events. 

\begin{mdframed}[backgroundcolor=gray!20] 
\textbf{\textit{Policy applications:}} Detecting anomalies in human behavior and mobility patterns could provide rapid assessment of damage inflicted to the built environment under data scarcity, and be applied for various downstream tasks including the preparation of real-time flood inundation maps and identifying dysfunctional mobile phone tower locations. \end{mdframed}

\subsection*{Epidemics}

Over the past decade, mobile phone location data have been utilized in modeling the outbreaks and spread of infectious diseases, including cholera, malaria, and Ebola. 
Many studies have used mobile phone CDR to extract mobility inter-regional fluxes, and have integrated such network dynamics with disease models to predict the spread of diseases and social dynamics \cite{fast2015modelling}. 
Moreover, the coronavirus disease 2019 (COVID-19) has spurred the use of mobile phone location data for disease modeling. 
In this section, we review the methods and case studies of the use of mobile phone location data for epidemic modeling. 

\subsubsection*{Mobility Network Estimation for Epidemiological Modeling}

The majority of the research have used mobile phone location data (mainly CDR) to extract the intra-regional mobility (origin destination) patterns, and integrates such insights into epidemiological models (e.g., SIR, SEIR models) to predict disease outbreaks. 
The seminal work on this topic performed by Wesolowski et al. used CDR from Kenya to quantify the importation routes that contribute to malaria epidemiology on regional spatial scales \cite{wesolowski2012quantifying}. 
The identification of the sources and sinks of imported infections due to human mobility showed significant potential in improving malaria control policies. 
Combined with rapid risk mapping, mobile phone location data based approaches could aid the design of targeted interventions to maximally reduce the number of cases exported to other regions while employing appropriate interventions to manage risk in places that import them \cite{tatem2014integrating}.  
A review and comparison of using survey based travel data and mobile phone data revealed that survey data produces lower estimates of travel, however, provided demographic information and motivations of travelers, which could be further utilized for modeling. 
On the other hand, mobile phone data provides a refined spatio-temporal description of travel patterns, although it lacks demographic information about the travelers \cite{wesolowski2014quantifying}.
Bengtsson et al. estimated the mobility network using movements of 2.9 million anonymous mobile phone users (CDR) in Haiti during the 2010 cholera outbreak. 
The prediction accuracy of the outbreak were compared with gravity model estimates, and it was shown that mobility networks generated from mobile phone data had comparable accuracy with gravity models, however, mobile phone data was advantageous since it required no model parameter calibration, unlike gravity models \cite{bengtsson2011improved}. 
Finger et al. used a mobile phone CDR dataset of over 150,000 users in Senegal to extract human mobility fluxes $Q_{ij}(t)$ $\forall i,j$ across regions $i$ and $j$, where $Q_{ij}(t)$ represents the community-level average fraction of time that users living in region $i$ spend in region $j$ during day $t$. 
By directly incorporating the mobility fluxes into a spatially explicit, dynamic epidemiological framework, they identified mass gatherings to be a key driver of the cholera outbreak  \cite{finger2016mobile}. 

\begin{figure}[t]
\centering
\includegraphics[width=.8\linewidth]{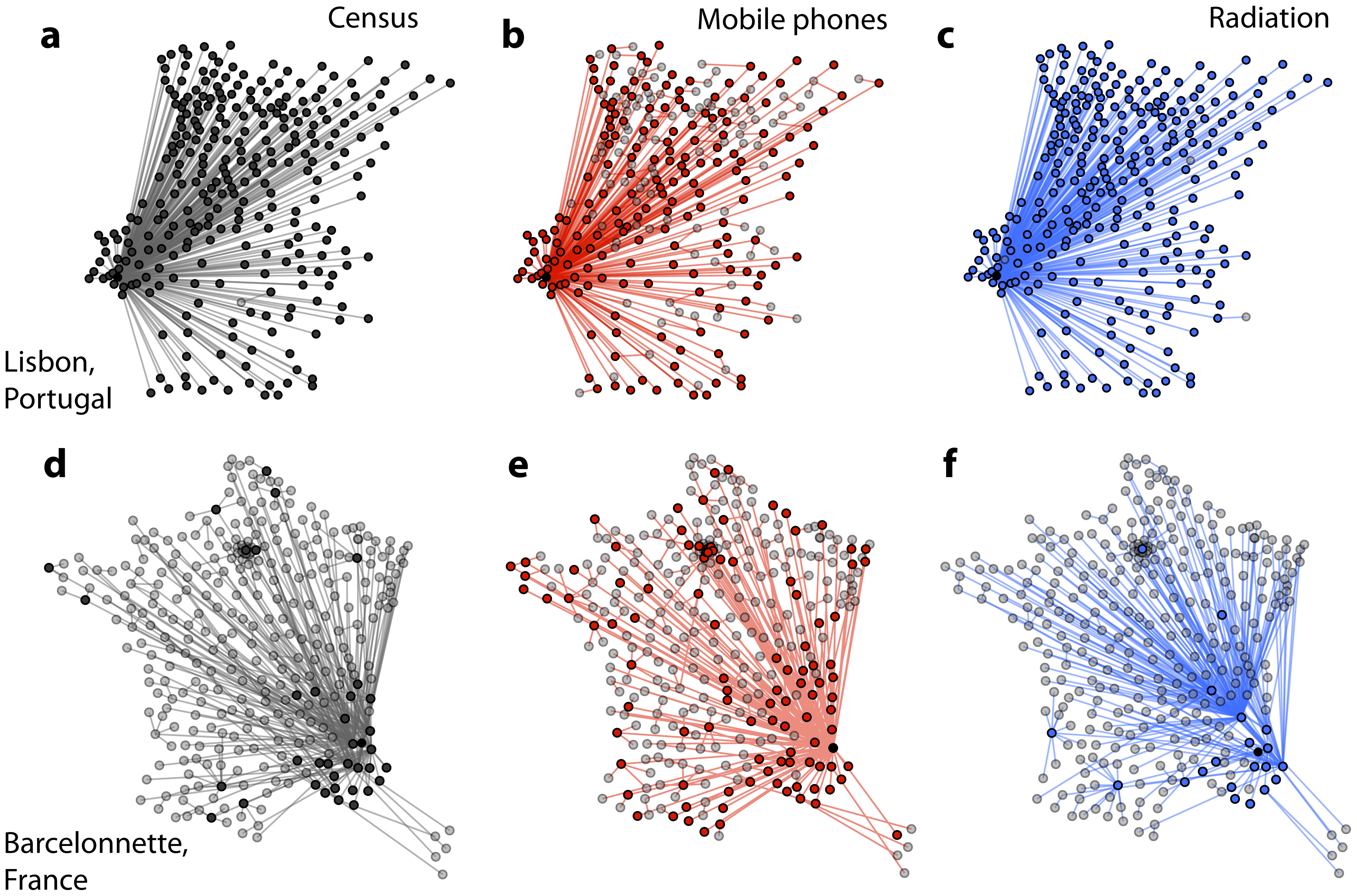}
\caption{\textbf{Epidemic invasion trees.} Full invasion trees for $R_0=3$ are shown for Portugal (top row) and France (bottom row) in the cases of the census network (a, d), the mobile phone network (b, e) and the radiation network. (Source: \cite{tizzoni2014use})}
\label{fig:epidemic}
\end{figure}

Similar studies have been conducted in other regions, on other types of diseases, with several improvements to the epidemic modeling methodologies. 
For example, Wesolowski et al. used seasonal fluctuations in travel patterns estimated from mobile phone data to characterize seasonal fluctuations in risk across Kenya for rubella disease \cite{wesolowski2015evaluating,wesolowski2015quantifying}. 
Moreover, seasonal asymmetric mobility patterns were used to refine the epidemiological models in Kenya, Namibia, and Pakistan \cite{wesolowski2017multinational}. 
Panigutti et al. assessed the stochasticity in the epidemic modeling outcomes, and showed that model estimates are become more adequate when epidemics spread between highly connected and heavily populated
locations \cite{panigutti2017assessing}. 
Similarly, Tizzoni et al. compared the use of mobile phone location data and census information in epidemiological models, and found that phone data matches the commuting patterns reported by census well but tends to overestimate the number of commuters, leading to a faster diffusion of simulated epidemics (shown in Figure \ref{fig:epidemic}) \cite{tizzoni2014use}.
Rubrichi et al. used mobile phone data and epidemiological models to evaluate the effects of various spatial-based targeted
disease mitigation strategies \cite{rubrichi2018comparison}. 
In other regional and disease contexts, Vogel et al. estimated the Ebola outbreak in Western Africa by using CDR in a simulation framework \cite{vogel2015mining}, Ihantamalala et al. estimated the sources and sinks of malaria parasites in Madagascar \cite{ihantamalala2018estimating}, and Kiang et al. improved forecasts of Dengue fever in Thailand by integrating human mobility data \cite{kiang2021incorporating}. 

During the coronavirus pandemic (COVID-19), we saw a rapid increase of the use of mobile phone smartphone GPS location data to estimate mobility networks for epidemiological modeling \cite{grantz2020use}. 
Schlosser et al. analyzed the structural changes in the mobility network during mobility restrictions in Germany, and found that long-distance travel trips were reduced disproportionately, enabling the flattening of the epidemic curve and delaying the spread to geographically distant regions \cite{schlosser2020covid19}. 
Lai et al. used mobility data-driven travel networks combined with an SEIR model to evaluate the effects of various non-pharmaceutical interventions on the spread of COVID-19 in China \cite{lai2020effect}. 

\begin{mdframed}[backgroundcolor=gray!20] 
\textbf{\textit{Policy applications:}} Estimating human mobility networks in spatial and temporal granularity can be used to not only understand migration patterns, but as crucial input for epidemiological models (e.g., SIR, SEIR models), which can be used to predict the outbreak of the diseases, and the effects of various policies (e.g., lockdown, inter-regional mobility restrictions) in containing the spread.
\end{mdframed}

\subsubsection*{Monitoring and Forecasting of Non-Pharmaceutical Intervention Effects}
While mobile phone location data has been shown to be an adequate data source to estimate the inter-regional mobility networks which are crucial inputs for epidemiological models, they can also be used to evaluate the effects of non-pharmaceutical interventions, including regional and national lockdowns and inter-regional travel restrictions, in restricting human behavior. 
Prior to the COVID-19 pandemic, Peak et al. used mobile phone CDR data to evaluate the effects of a lockdown in Sierra Leone during the Ebola epidemic \cite{peak2018population}. 
As many countries adopted non-pharmaceutical interventions (NPIs) during the COVID-19 pandemic, mobile phone location data (mainly GPS data) were used to evaluate the effects of such orders \cite{buckee2020aggregated}. 
Researchers from academia, industry, and government agencies (e.g., \cite{ukkusuri2020non}) have utilized large-scale mobility datasets to estimate the effectiveness of control measures in various countries.
Such analyses were conducted and often frequently updated to monitor mobility reduction situations \cite{kishore2020measuring}. 
Kraemer et al. used mobile phone-generated mobility data from Wuhan and detailed case data including travel history to show that especially during the early stages of the outbreak, the spatial distribution of the COVID-19 cases were explained well by mobility data \cite{kraemer2020effect}. 
Pepe et al. quantified three different aggregated mobility metrics (origin-destination movements between provinces, radius of gyration, and average degree of a spatial proximity network) during the lockdown in Italy using mobile phone location data \cite{pepe2020covid,bonato2020mobile}.

\begin{figure}[t]
\centering
\includegraphics[width=.9\linewidth]{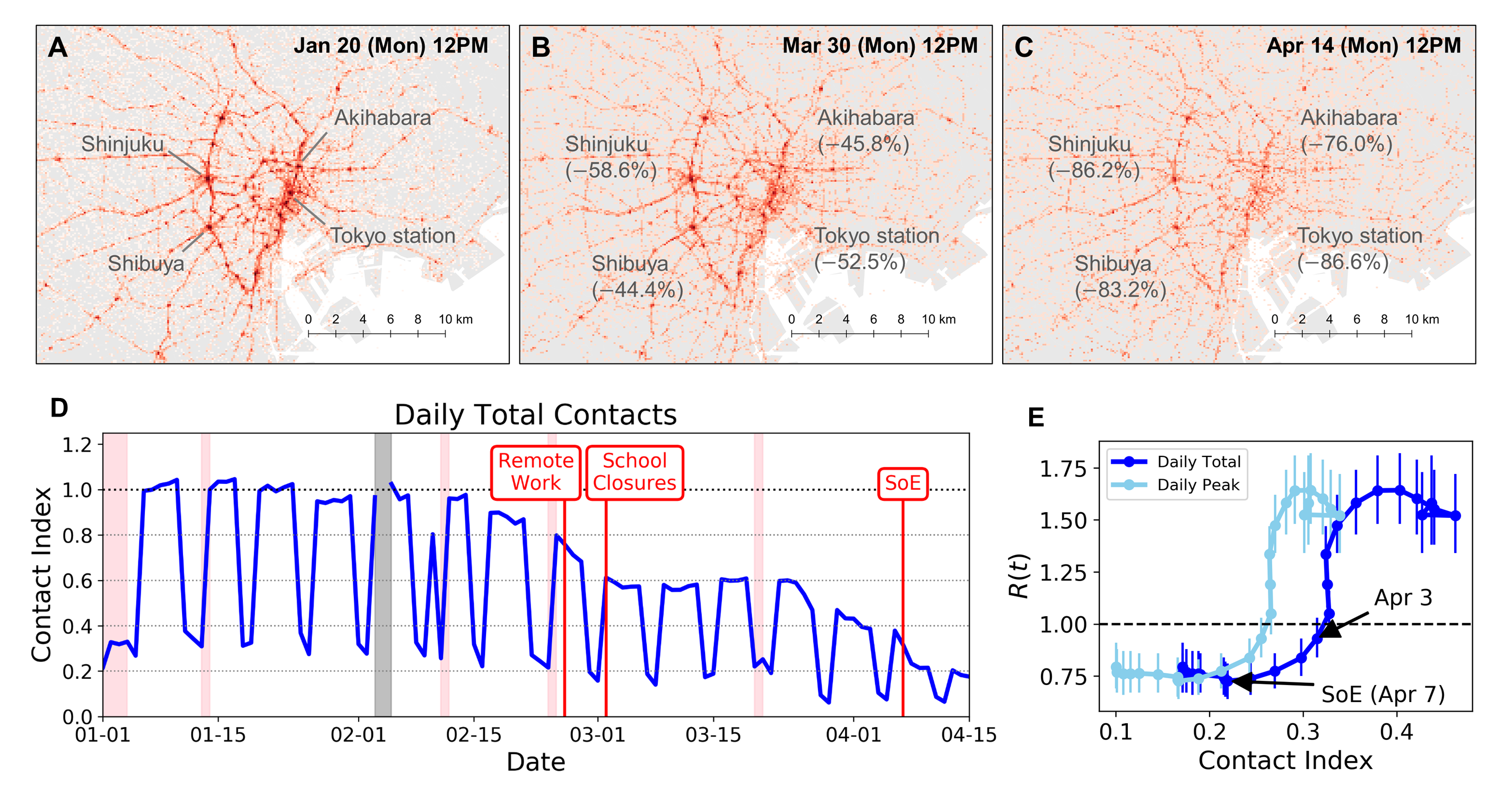}
\caption{\textbf{Macroscopic mobility dynamics.} (A)-(C) show the population distributions on 3 different dates at same times (12PM), each on the same day of week (Mondays). We observe substantial decrease in the population density at stations and cities. (D) shows the amount of contacts an individual potentially encounters outside home for each time period. (E) shows the non-linear relationship between the mobility metrics and $R(t)$. (Source: \cite{yabe2020non})}
\label{fig:timeline}
\end{figure}

In our past work in Japan as shown in Figure \ref{fig:timeline}, human mobility metric including the social contact index was quantified before, during, and after non-compulsory lockdowns \cite{yabe2020non}. 
Analysis showed that even after non-compulsory orders, mobility significantly dropped (70\% reduction) and the effective reproduction number had decreased to below 1. 
Such analysis has been conducted in the United States as well, assessing the effects of state-level interventions on mobility reduction \cite{wellenius2020impacts}, significant geographical variations in social distancing metrics \cite{gao2020mapping}, and income inequality in social distancing \cite{verma2021mobility}. 
Similar studies on mobility monitoring during non-pharmaceutical interventions were performed in Sweden \cite{dahlberg2020effects}, the United Kingdom \cite{santana2020analysis}, Italy \cite{cintia2020relationship}, France \cite{pullano2020evaluating}, Spain \cite{orro2020impact}, Switzerland \cite{molloy2021observed}, Finland \cite{willberg2021escaping}, Taiwan \cite{chang2021variation}, and Hong Kong \cite{zhang2021changes}.

In addition to monitoring the effects of non-pharmaceutical interventions, there has been an increasing number of studies focusing on forecasting and providing early warning of outbreaks.
Kogan et al. used multiple sources of data including mobile phone location data, social media data, and web search data, to prodive early warning signals of COVID-19 outbreaks.
The study showed that combining disparate health and behavioral data may help identify disease activity changes weeks before observation using traditional epidemiological monitoring \cite{kogan2021early}.
Similarly, Yabe et al. used mobility data and web search data provided by Yahoo Japan Corporation to develop risk indexes for microscopic geographical areas, and showed that such metrics could predict local outbreaks two weeks beforehand \cite{yabe2020early}.  
Chang et al. integrated human mobility network data into a metapopulation SEIR model to simulate the spread of COVID-19, and identified specific points-of-interest which are if closed, could be effective in suppressing the disease spread \cite{chang2021mobility}. 
The use of mobile phone location data has become prevalent in the field of economics, for example, Chetty et al. developed a platform to track the impacts of COVID-19 on businesses and communities in real-time, using various types of data including Google Mobility Report data \cite{chetty2020real}.

\begin{mdframed}[backgroundcolor=gray!20] 
\textbf{\textit{Policy applications:}} Quantifying various human mobility metrics (e.g., stay-at-home rates, average travel distance, social co-location index) in near-real time, in high spatial and temporal granularity, can be used to assess the effects of various non-pharmaceutical interventions on human behavior.
\end{mdframed}

\section*{Discussion}

\subsection*{Opportunities}

\subsubsection*{Increasing Availability of Data Products}

As reviewed in Section 3, many studies have already utilized the various kinds of mobile phone location data for disaster management. 
However, these were often enabled by direct partnerships or collaborations between researchers and private companies who own the data, making the data extremely difficult to access for researchers outside the agreement.
Due to the increased attention and interest on mobile phone location data during the COVID-19 pandemic, there has been several notable efforts where mobile phone location data, in their anonymized forms, are being made openly available for the public use. 
For example, the PlaceKey community (\url{https://www.placekey.io/}) have contributed to this effort by providing a semi-open platform where researchers can freely access aggregated mobile phone location data for analysis. 
The data are spatially and temporally aggregated to point-of-interests, and also made sure that a substantial small number of visit counts are masked, so that the individual users are unidentifiable. 
There are cases where researchers have led the efforts in anonymizing the data and making the mobility data open source. 
The team of researchers from The Robert Koch Institute and Humboldt University of Berlin have developed a dataset which contains mobility data collected from mobile phones in Germany during the first half of 2020 (January-July), and mobility data from March 2019, which can be used to study changes in mobility during the COVID-19 pandemic in 2020 (\url{https://www.covid-19-mobility.org/}).

In addition to these efforts, various organizations including major tech firms have made significant contributions in publishing aggregate statistics of mobility (e.g., social distancing, travel distance) during the COVID-19 for various regions around the world. 
The Google COVID-19 Community Mobility Reports, which contained the time series data of travelled distance in various cities around the world, was used by practitioners to monitor the effects of non-pharmaceutical policies on mobility restrictions \cite{aktay2020google}. 
A similar report on mobility patterns was also issued by Apple \cite{cot2021mining}.
Camber Systems developed the county-level social distancing tracker based on aggregated and anonymous location data to understand how populations are engaging in social distancing over time  (\url{https://covid19.cambersystems.com/}). 
The COVID-19 Mobility Data Network (CMDN) is a network of infectious disease epidemiologists at universities working with technology companies to use aggregated mobility data to support the COVID-19 response.
The CMDN developed the Facebook Data for Good Mobility Dashboard, which  visualizes the aggregate mobility trends, computed from Facebook mobility data, at the regional levels for various countries around the world (\url{https://visualization.covid19mobility.org/}).

\subsubsection*{Data for Development}
With the availability of various types of novel datasets including social media data, mobile phone location data (call detail records, GPS), web search query data, and satellite imagery data, there has been significant efforts to utilize big data analytics for tackling challenges in development \cite{hilbert2016big}. 
Several open data challenges have been initiated by collaborations between academia and industry data providers, such as the \textit{Data4Development Challenge} held by Orange, which provided mobile phone data from Ivory Coast for analysis \cite{blondel2012data}. 
Large tech firms, including Google, Facebook, Apple, and Microsoft, have all boosted their efforts in utilizing the enormous amount of collected data for development and disaster management. 
Google.org, the is the charitable arm of Google, has committed roughly US\$100 million in investments and grants to nonprofits annually to tackle various issues including disaster response, improving accessibility to education, and more recently, recovering from COVID-19 impacts (\url{https://www.google.org/}).  
International agencies have also accelerated their engagement in utilizing such big data sources for development projects.
The World Bank has initiated the Development Data Partnership (\url{https://datapartnership.org/}), which is a partnership between international organizations and companies, created to facilitate the use of third-party data in research and international development.
The Partnership includes more than 20 private companies, including location intelligence companies such as Google, Cuebiq, Safegraph, and CARTO, and social media companies including Twitter and Facebook.
To assist the utilization of these datasets, recently, the Global Facility for Disaster Reduction and Recovery (GFDRR) - a partnership hosted within the World Bank - has undertaken efforts on using GPS location data collected from smartphones to analyze post-disaster population displacement for disaster relief and urban planning policy making.
GFDRR has published working papers and publications on several case studies using smartphone location data accessed through the Development Data Partnership initiative, including the population displacement patterns and income inequality in Mexico City after the Puebla Earthquake \cite{gfdrrworkingpaper} and socioeconomic gaps in mobility reduction during the COVID-19 pandemic in Colombia, Mexico, and Indonesia \cite{fraiberger2020uncovering}.

\subsubsection*{Open Source Toolkits for Mobility Analytics}

To assist policy makers and non-data experts to leverage the increasing availability of mobile phone location datasets, there has also been several efforts to develop open source toolkits for mobility data analytics. 
\texttt{scikit-mobility} is a Python-based library that enables various operations and analyses on large-scale mobility data \cite{pappalardo2019scikitmobility}. 
Compared to previous Python based mobility analysis libraries such as \texttt{Bandicoot} \cite{de2016bandicoot} and \texttt{movingpandas} \cite{graser2019movingpandas}, \texttt{scikit-mobility} is most comprehensive, containing functions for pre-processing, stop detection, computation of mobility metrics (e.g., displacements, characteristic distance, origin-destination matrix), trajectory synthesis, visualizations, and privacy risk quantification.
There exists several libraries to conduct trajectory analysis in the R ecosystem, however, none of the libraries are optimized for human mobility data, thus lacks functions for generating synthetic trajectories and producing advanced visualizations (for a review, see \cite{joo2020navigating}).
\texttt{OSMnx} is a powerful library for acquiring, constructing, analyzing, and visualizing complex street networks from OpenStreetMap \cite{boeing2017osmnx}. 
In combination with human mobility data, \texttt{OSMnx} enables users to perform various spatial analysis including route estimation and point-of-interest visit estimation.
More recently, the GFDRR developed an open-source location data analytics toolkit in Python \texttt{MobilKit} in collaboration with Purdue University and MindEarth (a non-profit based in Switzerland \url{https://www.mindearth.org/}), which extends the functions in \texttt{scikit-mobility} to conduct post-disaster mobility analysis (\url{https://github.com/GFDRR/mobility_analysis}).
To enable non-experts to use the softwares, the codes are optimized using Dask \cite{rocklin2015dask} for parallel computing, so that analysis on massive mobility datasets can be conducted under constrained resources, on local laptop or desktop computers.  

\subsection*{Challenges}\label{challenges}

Despite the enormous opportunities in using mobile phone location data for disaster management as reviewed in the previous sections, the rise of novel mobile phone location data produced by location intelligence companies and the wide spread use of the data by various stakeholders, pose new challenges in utilizing the dataset in an inclusive, transparent, and sustainable manner. 
Here, we touch upon the main key challenges that we face in the usage of mobile phone location data, related to assuring the data quality, governance, and open research directions in developing advanced analysis techniques.  

\subsubsection*{Understanding the Data Generative Process}

One of the key drawbacks of using the more recently available smartphone GPS location data is the lack of our understanding in how these data are collected and processed. 
Several studies have conducted investigations on the representativeness of these datasets (e.g., \cite{yabe2020understanding}) using raw data, by quantifying the correlation between the number of mobile phone users estimated to be living in each geographical region, and the census population information. 
This metric, however, is far from comprehensive, and we have pressing demand for a more thorough investigation on various aspects of socio-demographic and socio-economic characteristics, and to ensure that the observation samples in the data are not biased towards a specific population group of wealth, region, ethnicity, gender, etc. 
This procedure becomes even more difficult when only aggregate information, such as the total number of daily users in a specific region or the daily number of visitors to a specific point-of-interest, are provided by the data providers. 
In addition to the uncertainties in the sample representativeness, the data collection procedure is not transparent. 
For example, some softwares and applications collect location data when the device detects substantial movement, therefore, only a very small number of points would be observed if the user stays at one location (e.g., home) during the entire day. 
Other algorithms collect location information in extremely high frequency (e.g., every minute), irrespective of the amount of movement. 
This is partly the reason why we observe such a large variance (i.e. truncated power law) in the number of observation points per user \cite{yabe2020understanding}. 
In the absence of methods and algorithms for correcting the bias in the data, the trustworthiness of the data products and analysis will be undermined. 
A more open discussion between data users -- researchers and practitioners -- and data providers to further understand the process of dataset generation, and a standardized way of quantifying and reporting the representativeness biases and the potential errors present within the dataset are essential for more inclusive, fair, and trustworthy data products for disaster response.


\subsubsection*{Data Governance}

As we experience an increase and universal accessibility to large scale mobile phone location data, the protection of personal privacy has never been more important \cite{nyt}. 
Previous studies have revealed that a very few number of data points could reveal the identity of the user with high accuracy, highlighting the importance of anonymization techniques \cite{de2013unique}.
Following such public concerns, data providers have started to provide processed data, aggregated by space and time. 
For example, the Disaster Maps data in the Facebook Data for Good program aggregates population density and flow into each day, into 6 kilometer size grid cells, and further applies spatial smoothing algorithms to anonymize the data. 
This process, although effective in anonymizing the data and protecting the users' privacy, comes with a price in the data granularity and uncertainties in the data quality, as explained in the previous Section. 
To address this issue and to balance out the data quality with privacy protection, the concept and techniques of differential privacy are gaining attention. 
Differential privacy is a criterion, which tools are devised to satisfy. 
It enables the collection, analysis, and sharing of statistical estimates using personal data while protecting the privacy of the individuals in the dataset \cite{wood2018differential}. 
Techniques such as differential privacy may serve as one baseline to ensure the safety of personal privacy, but we are still amidst the search for a holistic framework that integrates technical solutions, ethical guidelines, and regulations on the use of mobile phone location data.

\subsubsection*{Translating Analysis into Disaster Risk Management Policy and Operations}
Mobility data has been shown to be effective in various applications that can be used for policy inputs, including conducting rapid post disaster damage and needs assessment, business disruptions and recovery monitoring, and dynamic population mapping. Key areas of opportunity include: (i) increasing situational awareness of emergency response managers through timely information on the number and geographic location of displaced persons; (ii) improving damage and needs assessments through quantification of foregone economic activity in business sectors; (iii) informing finance and policy support for post-disaster recovery efforts by quantifying business recovery rates in affected districts.
While there has been many successful cases of translating the mobility data analysis into policy decision making, such as the population displacement maps after the Gorkha Earthquake \cite{wilson2016rapid}, there is still a limit to the number of organizations and research groups that are capable of conducting such an end-to-end, analysis-to-policy translation. 
As many regions face an increasing likelihood of experiencing disaster events due to urbanization and climate change, there is a pressing demand for expanding these mobility data-driven solutions across regions and disaster events. 
One attempt to localize these data-driven solutions is to develop open source toolkits (as introduced in Section 4.1.3) to increase the capacity of local stakeholders and data scientists to conduct such analysis. 
In addition to the analytics tools, stakeholders require an effective scheme to share experiences, knowledge, and know-how across different regions and stakeholders. To foster strong uptake of insights derived from human mobility data, further methodological research is needed to address key challenges such as quantifying the representativeness and socio-economic bias of human mobility datasets; and accounting for impact of network outages during disaster events on observed population numbers.
Exploring a way to expand the mobility data analytics into various local contexts is a critical operational challenge.

\subsection*{Future Research Directions}


\subsubsection*{Cross-Comparative Analysis across Events}
Mobile phone location data, with its global coverage and spatio-temporal scale in data size, allows us to conduct cross comparisons across locations, disaster types and time scales, as shown in previous studies (e.g., \cite{yabe2018cross,yabe2020understanding}). 
Comparing the response and recovery dynamics across different disaster events across regions, allows us to extract essential dynamics that govern the disaster recovery process, as demonstrated in Yabe et al., where general patterns of recovery of population displacement (i.e., negative exponential decay) were discovered \cite{yabe2020understanding}. 
In addition, such parsimonious models allow us to show and explain the differences/variability that exist across the regions, using socio-demographic and -economic factors.
Such type of transferable and universal models are much needed in the disaster science literature. 
Using such insights, we are able to build parsimonious models of disaster response and recovery, which were difficult to do before using conventional household survey data.

\subsubsection*{Modeling the Disaster Recovery Process}

As pointed out in Section 3.1.2, although we have a large collection of case studies that conduct displacement analysis using mobile phone location data, the literature is still limited in studies that perform analysis and modeling of the long-term recovery process after natural hazards. 
This is partly due to the lack of availability in long term data (over 6 months) in the same region and data provider. 
As a result, one limitation in natural hazard response and recovery process modeling is the lack of a standardized parsimonious model that captures the dynamics of population movement and recovery, similar to what the SIR, SEIR models achieve in epidemiological modeling. 
Mobile phone location data, either independently or by fusing with other data types, allows us to model correlations and interdependencies across various systems that compose cities.
Modeling these interdependencies will allow us to build dynamic and causal models that show how the social, built environment and economic forces contribute to the response and recovery of communities after disasters.
Recently, a system dynamics modeling approach that captures the interdependent dynamics between social and technical systems has been proposed and tested using the case study of recovery after Hurricane Maria in Puerto Rico \cite{yabe2021resilience}. 
Despite its capability in replicating the recovery process and understanding the system interdependencies that play a role in disaster recovery, we still lack parsimonious models that capture the various aspects of disaster recovery including population migration.

\subsubsection*{Fusion with Other Data Sources}
While we have seen a rapid increase of the usage of mobile phone location data, there are several other types of data that have been used frequently in disaster management, including satellite imagery (for a review article, see \cite{joyce2009review}) and social media data (for review articles, see \cite{martinez2018twitter,muniz2020social}). 
Satellite imagery, despite its low frequency of data collection, enables the observation of damages to the natural and built environments in a detailed spatial scale. 
On the other hand, social media data contains rich information on the peoples' opinions, ideas, and sentiments at a high temporal granularity. 
Moreover, combining mobile phone location data with household surveys could allow us to analyze both the post-disaster mobility patterns as well as the motivations behind such behavior. 
More recently, credit card transaction data has become more available for research purposes (e.g., \cite{di2018sequences}). 
Using credit card data, we are able to understand the economic impacts of disasters and epidemics at a spatially and temporally granular level. 
Combining these datasets with mobile phone location data and human mobility analytics (e.g., application in poverty estimation \cite{steele2017mapping}) could enable a more holistic understanding of the social, physical, and economic dimensions of the disaster response and recovery dynamics.

\section*{Conclusions}

Due to rapid urbanization, climate change, and complex interactions with various social, economic, and political factors, the risks of disasters -- natural hazards and pandemics -- are continuing the increase across the globe. 
The comprehensive literature review of research and efforts that have used mobile phone location data for (natural and pandemic) disaster management throughout the past decade has shown that such data enables the implementation of various rapid, high-precision, large-scale approaches to assist disaster management (response, recovery, and preparation), compared to conventional approaches using household surveys.
More specifically, applications in natural hazard response include population displacement and recovery analytics, quantifying economic disruptions, and inferring the physical damage inflicted to critical infrastructure systems through behavioral changes. 
Dynamic and high-resolution origin destination matrices are critical inputs for epidemiological models, and have already been applied to predict the spread of a wide range of communicable diseases.  
The COVID-19 pandemic spurred the use of mobile phone location data for pandemic disaster response, showcasing the usefulness of the data in pandemic response and recovery.

With both the increase in demand and supply of location-based intelligence platforms and applications both from public and private entities, we anticipate that the availability of location and human mobility datasets to continue its increasing trend. 
The review of available data products, data-sharing ecosystems (e.g., Development Data Partnership of The World Bank), and open source toolkits for the analysis of human mobility data for disaster response and recovery applications highlighted the growth of the data-ecosystem around human mobility data. 
Human mobility data, more specifically mobile phone location data, holds immense opportunities for a more efficient, inclusive, and adaptive disaster response and recovery. 
To standardize and operationalize the usage of human mobility data in disaster response and recovery, however, requires progress in various interdisciplinary aspects; technical methods for equitable and fair analysis and policy translation, and data governance protocols for safe and responsible usage of mobility data.

Moving forward, we face challenges in both technical and governance aspects of mobile phone location data. 
Despite the increase in studies using mobile phone data, there are still significant research gaps in analysis methodologies and techniques for longer-term disaster recovery modeling using mobility data. 
Although human mobility data provided by location intelligence firms have numerous advantages over call detail records, the data generative process is often undisclosed and inaccessible from the user-end. 
To avoid unintended and undesirable consequences of using biased and unrepresentative data, more preliminary analysis on this topic is urgently needed.  
The wide usage of mobility data during the COVID-19 crisis has increased the awareness of its inherent risks to privacy, not just within the research community but among the entire society. 
Further discussions are required around identifying the appropriate methods and frameworks involving various stakeholders -- researchers, tech companies and data providers, citizens, and policy makers -- to address issues around data quality and data governance. 
Moreover, despite the increasing number of efforts in linking mobility data analysis products to policy making driven by multiple stakeholder levels, we are still amidst the quest of searching for an optimal stakeholder engagement framework that enables effective policy translation.

Reviewing the methods and applications of mobile phone location data shed light on the opportunities, challenges, and most importantly, several key questions and future research directions, namely 1) going deeper, 2) horizontal scaling, and 3) heterogeneous data mixing.
Going deeper into the data itself and developing a more holistic and systematic understanding of the data generative process is key for achieving a more equitable usage of mobile phone location data, especially for policy applications.
Scaling the analysis horizontally by making a leap from individual case studies focusing on specific regions and disaster events towards a global analysis of different events is key to generating universal and transferable insights. 
Eventually, as a research community, we need to aim to build a standardized model of disaster displacement and recovery, similar to the SIR and SEIR models in epidemiology.
Technical methods for the fusion of mobile phone location data with other heterogeneous datasets including satellite imagery, household surveys, and social media data is a relatively untouched area of research. 
We believe that conducting this review at this timing, where COVID-19 has spurred the use of mobile phone location data, is particularly valuable in organizing the expanding literature, and contributes to the field by setting key agenda items and directions for the next decade of research using mobile phone location data. 

\bibliography{sample}



\section*{Acknowledgements}
The work of T.Y. and S. V. U. is partly funded by NSF Grant No. 1638311 CRISP Type 2/Collaborative Research: Critical Transitions in the Resilience and Recovery of Interdependent Social and Physical Networks.

\section*{Author contributions statement}
All authors wrote and reviewed the manuscript. 

\section*{Additional information}
The authors declare no conflict of interest.




\end{document}